# Chebyshev Methods for Ultra-efficient Risk Calculations

Spectral Decomposition Applications in Risk Management

**Mariano Zeron Medina Laris[1], Ignacio Ruiz[2]**

April 2018

**Abstract**

*Financial institutions now face the important challenge of having to do multiple portfolio revaluations for their risk computation. The list is almost endless: from XVAs to FRTB, stress testing programs, etc. These computations require from several hundred up to a few million revaluations. The cost of implementing these calculations via a "brute-force" full revaluation is enormous. There is now a strong demand in the industry for algorithmic solutions to the challenge. In this paper we show a solution based on Chebyshev interpolation techniques. It is based on the demonstrated fact that those interpolants show exponential convergence for the vast majority of pricing functions that an institution has. In this paper we elaborate on the theory behind it and extend those techniques to any dimensionality. We then approach the problem from a practical standpoint, illustrating how it can be applied to many of the challenges the industry is currently facing. We show that the computational effort of many current risk calculations can be decreased orders of magnitude with the proposed techniques, without compromising accuracy. Illustrative examples include XVAs and IMM on exotics, XVA sensitivities, Initial Margin Simulations, IMA-FRTB and AAD.*

[1] MoCaX Intelligence. London, UK. m.zeron@iruiztechnologies.com
[2] MoCaX Intelligence. London, UK. i.ruiz@iruiztechnologies.com

# Introduction

There is a wide range of situations in finance where pricing functions must be evaluated many times for risk computation. Examples include all the XVAs for pricing, Counterparty Credit Risk capital via IMM (IMM-CCR), Market Risk VaR and stress-VaR (sVaR), the upcoming FRTB advance IMA approach (IMA-FRTB), stress-testing frameworks, CCAR, etc. They all require from a few hundred to a few million portfolio revaluations in each computation.

The complexity of pricing functions has been increasing over the last few decades; from vanilla European options and swaps to all the Bermudan-swaption family of derivatives, exotic FX and equity products, credit derivatives, commodity products, etc.

As a result, the computational effort (i.e. time and hardware cost) taken to perform such risk evaluations has become a problem. For example, tier-one banks are now estimating the hardware costs of being able to do the IMA-FRTB computation via a "brute-force" hardware approach (i.e. building a computing grid powerful enough to do the calculation in true full-revaluation mode, calling the Front Office pricing functions as often as needed). These costs tend to range from $50 million to around $200 million, and this is considering only the compulsory capital calculations; live intra-day calculations that would enable the right business and trading decision with an ex-ante analysis of impact in FRTB capital would be additional.

Solutions to this important issue lie at the core of being able to face the demands imposed by both industry and regulators. If this computational problem is not solved intelligently, many business units and trading desks will become uneconomical and will have to close. On the positive side, those banks that manage to solve this problem will have a competitive advantage.

In this paper, we introduce an optimal method which helps tackle this problem.

# Mathematical Framework

In this section we establish the mathematical framework which will later be applied in the context of Risk Calculation Engines.

## Orthogonal Expansions

There are a number of spectral decomposition methods that are based on projecting functions onto the subspace generated by a series of basis functions. Depending on what basis is chosen, and its dimensionality, we can obtain optimal solutions.

Central to this is the use of Chebyshev polynomials and Chebyshev interpolants. In many cases, we are able to construct approximators which can be efficiently evaluated. This means that, with an extremely low number of evaluations of the original function, we are able to generate an approximating function that is extremely accurate.

Putting this into practice, we are looking for a replica of a pricing function, which can be built based on the value of the pricing function at a few points, which is as accurate as required and which is extremely fast to evaluate. Such a replica can then replace the original pricing function in the millions of evaluations required in a risk calculation, thus completing the calculation in a small fraction of the time, without sacrificing any significant accuracy.

Polynomial functions are a well-understood family of functions used in various areas of mathematics. One of their uses has been to approximate arbitrary continuous functions. Going back as far as 1885 we find evidence of their suitability for approximation (Weierstrass, 1885)

**Theorem 1**. *Weierstrass approximation Theorem*. Let $f$ be a continuous function on $[-1,1]$, and let $\varepsilon$ be an arbitrary real value greater than zero. Then there is a polynomial $p$ such that

$$||f - p||_\infty \leq \varepsilon$$

This result, however, does not give a recipe for how to find, given a continuous function $f$, a suitable approximating polynomial.

There are two things to bear in mind when deciding what polynomial $p$ to use as an approximator. On the one hand, we want accuracy. On the other, particularly in applications like the one we are interested in, we want a computer to evaluate the approximator as efficiently as possible.

Often, these two aims pull in different directions. Generally, the more accuracy we need, the more complex the evaluation of the approximator becomes. Let us express this more concretely.

Denote the space of polynomials by $\mathcal{P}$. Consider its canonical filtration

$$\mathcal{P} = \bigcup_{n \in \mathbb{N}} \mathcal{P}_n$$

where $\mathcal{P}_n$ is the space of polynomials of degree at most $n$. The higher the $n$, the more polynomials available to choose from. If we cannot find a polynomial $p$ in $\mathcal{P}_n$ that approximates our function $f$ sufficiently well, we need to look for polynomials in $\mathcal{P}_m$ where $m > n$. The higher the $m$, the higher the degrees of the polynomials. However, the higher the degree $m$, the longer it takes to evaluate.

We want a systematic way to find, for low values of $n$, polynomials in $\mathcal{P}_n$ that approximate $f$ to the degree we need. This is central to the applications of these methods, as industrialised risk computation settings require systematic solutions to implement and maintain.

It is in this sense that the theory of Chebyshev interpolants provides an extraordinary balance between the two as will be seen in the following sections.

## Chebyshev Series

So far, we have spoken of using polynomials to approximate any given continuous function. However, the results that constitute the core of the theory of Chebyshev approximation are valid for continuous functions that enjoy a certain degree of smoothness. Though in principle restrictive, this turns out to be particularly useful in our context, as pricing functions in finance are often not only piecewise continuous but piecewise differentiable and, indeed, in most cases, piecewise analytic.

In the following sections, unless otherwise specified, we work with functions of one dimension defined on the closed bounded interval $[-1,1]$. If the function in question $f$ is defined on a generic interval $[a, b]$, we pre-compose it by a scaling function that takes the interval $[-1,1]$ to $[a, b]$. By doing this we obtain a function $g$ defined on $[-1,1]$. Every result we state from now on valid for $g$, will also be valid for $f$. Therefore, we assume, with no loss of generality, that the domain of the functions we are interested in is the interval $[-1,1]$.

**Definition 2.** A real-valued function $f: \mathbb{R} \to \mathbb{R}$ is called Lipschitz continuous if there exists a positive real constant $K$ such that, for all real $x_1$ and $x_2$,

$$|f(x_1) - f(x_2)| \leq K|x_1 - x_2|$$

**Definition 3.** The $k$-th Chebyshev polynomial is defined as

$$T_k(x) = \cos(k\theta),$$

where

$$\theta = \cos^{-1}(x).$$

For alternative definitions of Chebyshev polynomials, as well as their main properties we refer to (Handscomb & Mason, 2003).

The following results are central in the theory of approximation with Chebyshev polynomials. Proof of the following Theorems may be found in (Handscomb & Mason, 2003) and (Trefethen, 2013).

**Theorem 4.** If $f$ is Lipschitz continuous on $[-1,1]$, it has a unique representation as a Chebyshev series,

$$f(x) = \sum_{k=0}^{\infty} a_k T_k(x)$$

which is absolutely and uniformly convergent. The coefficients are given for $k \geq 1$ by the formula

$$a_k = \frac{2}{\pi} \int_{-1}^{1} \frac{f(x) T_k(x)}{\sqrt{1-x^2}} dx$$

If $k = 0$, the above formula is divided by 2.

## Truncated Chebyshev Series

For a fixed natural number $n$, a good candidate to approximate Lipschitz continuous functions is the truncation of the Chebyshev series up to its $n$-th degree

$$f_n(x) = \sum_{k=0}^{n} a_k T_k(x)$$

We call this the Chebyshev projection of degree $n$.

So far, we have convergence of Chebyshev series for Lipschitz continuous functions. This is a good starting point, but we need to make sure the desired degree of accuracy is obtained with the smallest possible $n$. Otherwise, the computational effort to evaluate $f_n$ could be too high for it to be useful in contexts where speed and precision are critical. Therefore, it is important to use polynomials with the smallest degree possible. As we will see, the Chebyshev series enjoys phenomenal convergence rates for a substantial class of functions. This ensures the projections we use have the smallest possible degree (i.e. $n$ is small).

If the function is not only Lipschitz continuous but differentiable of order $m$, we have

**Theorem 5**. For an integer $m \geq 0$, let $f$ and its derivatives up to and including $f^{(m-1)}$ be absolutely continuous on $[-1,1]$. Furthermore, suppose the $m$-th derivative $f^m$ is of bounded variation $V$. Then for any $n > m$, its Chebyshev projections satisfy

$$||f - f_n||_\infty \leq \frac{2V}{\pi m (n-m)^m}$$

A function $f$ is *analytic* if for every point $x$ in its domain, the Taylor Series of $f$ at $x$ exists and converges to the value $f(x)$. When a function is analytic on $[-1,1]$ it can be analytically extended to a neighbourhood in the complex plane around $[-1,1]$. For the following Theorem, the regions of interest are ellipses with foci at $-1$ and $1$, called Bernstein ellipses. The bigger the ellipses, the faster the convergence. The following Theorem, which strengthens the result of Theorem 5, first appeared in (Bernstein, 1912).

**Theorem 6**. Let $f$ be an analytic function on the interval $[-1,1]$. Consider its analytical continuation to the open Bernstein ellipse $E_\rho$, where it satisfies $|f(x)| \leq M$, for some $M$. Then for each $n \geq 0$

$$||f - f_n||_\infty \leq \frac{2M\rho^{-n}}{\rho - 1}$$

where $f_n$ is the Chebyshev projection of degree $n$.

As mentioned before, a good portion (if not all) of the functions used in finance (e.g. pricing functions) satisfy strong degrees of differentiability or are indeed analytic[3]. For functions with the latter property, projections of the Chebyshev series with small degree often give very good degrees of approximation because of the exponential convergence of $f_n$ to $f$.

### Interpolants

As is evident from the Theorems in the previous section, Chebyshev projections offer a very attractive way of approximating piecewise analytic (or even differentiable) functions. However, evaluating such projections may be tricky as the coefficients in Theorem 4 involve solving the integral in question.

---

[3] In all but a finite set of points of their domain. For example, barriers on barrier options or payment dates on interest rate swaps. In those cases, the methods described can be extended to piecewise functions without loss of generality.

However, there are polynomial interpolants which enjoy the same convergence rate as the ones specified in Theorems 4, 5, 6, and have expressions that are much easier to find and evaluate. In order to introduce such polynomial interpolants, we need the following.

**Definition 7**. The Chebyshev points associated with the natural number $n$ are the real part of the $n-$th roots of unity $z_j$

$$x_j = Re z_j = \frac{1}{2}(z_j + z_j^{-1}), \quad 0 \leq j \leq n$$

Equivalently, Chebyshev points can be defined as

$$x_j = \cos(j\pi/n), 0 \leq j \leq n.$$

These points are the projections of equidistant points in the upper half of the unitary circle in the complex plane onto the real line.

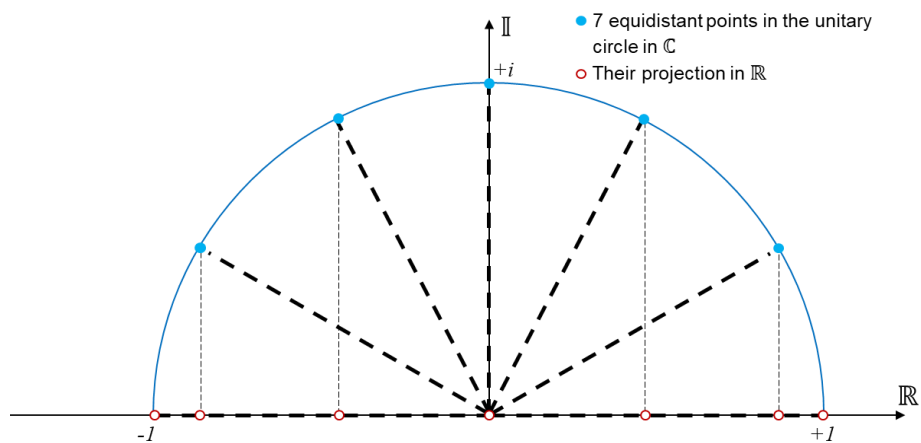

For more details on Chebyshev points, properties, and generalisations we refer to (Handscomb & Mason, 2003).

Let $f$ be a function and let $x_0, \ldots, x_n$ be points in the domain of $f$. Denote by $v_0, \ldots, v_n$ the values of $x_0, \ldots, x_n$ under $f$. We call the points $x_0, \ldots, x_n$ Anchor points and $v_0, \ldots, v_n$ the values of $f$ on the Anchor points. An interpolant to $f$ on such points is defined as a function $p$ such that $p(x_i) = f(x_i)$, for all $i, \ 0 \leq i \leq n$. If $p$ is a polynomial, then we say $p$ is a polynomial interpolant to $f$ on the Anchor points $x_0, \ldots, x_n$. It is a well-known result that for every function $f$ and every $n+1$ points such as the ones above, there is a unique polynomial interpolant $p$ of degree at most $n$.

Polynomial interpolants have a bad reputation as approximators in numerical computing. This bad reputation is partly due to a result by Faber in 1914 (Faber, 1914), which says that for the class of continuous functions there is no polynomial interpolation scheme, regardless of the way in which the points are distributed, that will ensure convergence. As a result of this theorem, polynomial interpolations have been the victim of an unfair negative reputation because, as soon as we restrict the functional space to *Lipschitz continuous functions,* we have guaranteed convergence of polynomial interpolations and, furthermore, as soon as we restrict ourselves to *analytic functions*, polynomial interpolations can be exponentially convergent, as we will see next, if the interpolation scheme is the right one. In addition, in practical risk calculation, most pricing functions that we work with are either analytic, or piecewise analytic in known segments.

It is crucial to stress the fact that the right interpolation scheme is central to the results presented in this paper. Indeed, if we do not work with the right interpolation scheme, even restricting the class of functions to analytic does not necessarily help matters. Interpolation on equidistant points, which is generally considered a sensible choice, will diverge even for functions with a high degree of smoothness, as the following example shows.

**Example 8**. A famous example was given by Runge in 1901 (Runge, 1901). He showed that equidistant interpolation not only diverges for the function

$$\frac{1}{(1+25x^2)}$$

but it diverges exponentially. This is known as the Runge phenomenon.

Polynomial interpolation on equidistant points has other problems too. Even when convergence is guaranteed in theory, rounding errors in floating point arithmetic may cause problems.

**Example 9**. Mathematically, polynomial interpolation on equidistant points converges for $exp(x)$. However, rounding errors in floating point arithmetic are amplified as the degree of the polynomial increases, making it diverge in practice.

## Chebyshev Interpolants

What is often missed is **that interpolation on sets of points different from equidistant can completely change the properties of polynomial interpolants**. In particular, polynomial interpolation on Chebyshev points has the same convergence properties as Chebyshev projections. Moreover, expressions of such polynomials are much easier and quicker to both find and evaluate as we describe next.

Denote the polynomial interpolant to a continuous function $f$ on the first $n+1$ Chebyshev points by $p_n$. Such a polynomial lies in $\mathcal{P}_n$, the space of polynomials of degree at most $n$. As such, it may be expressed as a linear combination of the first $n+1$ Chebyshev polynomials (indeed, as the linear combination of any basis for $\mathcal{P}_n$)

$$p_n(x) = \sum_{k=0}^{n} c_k T_k(x), \quad (1)$$

Expressing the Chebyshev interpolant in terms of Chebyshev polynomials has the following advantage. As it is shown in (Ahmed & Fisher, 1970), one can compute the coefficients $c_k$ from the values of $f$ on Chebyshev points, by using the Fast Fourier Transform. This has important consequences as the Fast Fourier Transform can be applied in $O(nlogn)$ operations. The expression in Equation 1 is not only easy and quick to find but can also be efficiently computed in a way that is numerically stable and unaffected by rounding errors in computers.

The reason why $p_n$ has such good convergence properties has to do with a phenomenon called aliasing. The full details of aliasing are beyond the scope of this paper. The following Theorem (Clenshaw & Curtis, 1960), however, sheds some light on the nature of this phenomenon and sets the stage for a very useful technique that helps control the error of Chebyshev interpolants in a very effective way.

**Theorem 10**. Let $f$ be Lipschitz continuous on $[-1,1]$ and let $p_n$ be its Chebyshev interpolant in $\mathcal{P}_n$, $n \geq 1$. Let $a_k$ and $c_k$ be the Chebyshev coefficients of $f_n$ and $p_n$, respectively. Then

$$c_0 = a_0 + a_{2n} + a_{4n} + \cdots, \qquad (2)$$

$$c_n = a_n + a_{3n} + a_{5n} + \cdots,$$

and for $1 \leq k \leq n - 1$,

$$c_k = a_k + (a_{k+2n} + a_{k+4n} + \ldots) + (a_{-k+2n} + a_{-k+4n} + \ldots) \quad (3)$$

An important consequence of Theorem 10 is the following. Theorem 4 says that the Chebyshev Series is uniformly and absolutely convergent. This implies that as $n$ tends to infinity, the absolute values of $a_n$ tend to zero. This, coupled with the equations in Theorem 10, imply that the greater the $n$, the smaller the terms $(a_{k+2n} + a_{k+4n} + \cdots)$ and $(a_{-k+2n} + a_{-k+4n} + \cdots)$, which means the terms $c_k$ and $a_k$ get closer and closer, for $1 \leq k \leq n - 1$.

On the one hand, the coefficients $a_n$ give us an idea of when we are close to converging. On the other, when we are close to converging, the coefficients $c_k$ approximate the coefficients $a_k$ ($1 \leq k \leq n - 1$). As the coefficients $c_k$ are much easier to compute than the coefficients $a_k$, we can use the former to identify potential convergence.

There is, of course, no guarantee that the series of Chebyshev interpolants has converged to the function if say, the last few coefficients $c_k$ are below a desired threshold. In practice, however, and specially for analytic functions, once the absolute value of the coefficients $c_k$ are below a small threshold (e.g $10^{-4}$) we can be pretty sure our Chebyshev interpolant will give us at most this error. This has practical implications as one can **control the error of the approximation without knowing anything about the function, apart from the values of the function on Chebyshev points and the fact that it is Lipschitz continuous.**

Mathematically, we can state the convergence properties for Chebyshev interpolants in the following manner. The following Theorems make use of Theorems 5,6 and 10. Details of the proofs, can be found in (Trefethen, 2013).

**Theorem 11**. Let $f$ satisfy the conditions of Theorem 5. Then for any $n > m$

$$||f - p_n||_\infty \leq \frac{4V}{\pi m (n - m)^m}$$

**Theorem 12**. Let $f$ satisfy the conditions of Theorem 6. Then for each $n \geq 0$

$$||f - p_n||_\infty \leq \frac{4M\rho^{-n}}{\rho - 1}$$

**Remark 13**. Looking at Theorems 11 and 12, it is easy to see why polynomial interpolation on Chebyshev points is a much more attractive method for approximation than others typically used. These theorems are extensions of Theorems 5 and 6 (that apply to the truncated Chebyshev series $f_n$) to Chebyshev polynomial interpolants $p_n$. Theorem 12 proves that the rate of convergence of such polynomial interpolant is also exponential. This property is central to the benefits this theory can bring to computation of risk in portfolios of financial derivatives.

However, polynomial interpolation on points other than Chebyshev (or similarly distributed) is ill-conditioned as is mentioned in Example 8. Even when convergence is guaranteed, rounding errors can create problems as Example 9 mentions.

Typical linear and spline interpolation have convergence properties for a wide collection of functions. However, the rate at which they converge is very slow. It is not exponential, even for analytic functions. In the context we are interested in applying these techniques, this translates into lots of calls to the original pricing function to achieve a minimum degree of accuracy, which translates into high pricing CPU time, high hardware costs, or both. This is obviously suboptimal when confronted with the optimal properties of Chebyshev interpolants.

In addition, with linear and spline interpolation we have limited or no control of the error of the approximation. However, as we have seen from Theorem 10, Chebyshev offers a most valuable ex-ante control of the error the approximation has.

Machine learning methods in general, even sophisticated ones, are good in regions where a lot of data is present but struggle to capture the values associated to areas of the domain where there is little data available. This sharply contrasts with Chebyshev approximation as convergence is proved with respect to the supremum norm, hence every point in the domain must satisfy the formulas in Theorems 11 and 12. In addition to not being able to capture tail events with accuracy, the calibration of machine learning methods can be tricky and may require calling the pricing function many more times than needed for the calculation in hand, reducing and sometimes killing the time savings we were after.

Taylor expansions, widely used in the industry for their speed, are usually only expanded to first or second order. Their accuracy can therefore not compete with the accuracy of Chebyshev interpolants.

As a counterpoint to the previous comments, we have the following Remarks.

**Remark 14**. Chebyshev interpolation compares very favourably even to optimal polynomial approximation. Denote by $p_n^*$ the polynomial in $\mathcal{P}_n$ that minimises $||f - q_n||_\infty$, the distance between $f$ and $\mathcal{P}_n$, where $q_n$ denotes a generic polynomial in $\mathcal{P}_n$. The polynomial $p_n^*$ is basically the best possible polynomial approximation to $f$ of degree less than or equal to $n$. Such polynomial, though in principle ideal for approximation purposes, is often difficult to compute. Not only are Chebyshev interpolants easy to compute, but as proved in (Ehlich & Zeller, 1966), if $p_n$ is the Chebyshev interpolant of degree $n$, $||f - p_n||_\infty$ cannot exceed $||f - p_n^*||_\infty$ by more than the factor $2 + (2/\pi)log(n + 1)$. Moreover, if $f$ is analytic, such factor is only 2.

**Remark 15**. Reducing a 1-dimensional function $f$ to an expression such as the one in Equation 1 has several advantages. The first and most obvious is being able to evaluate it in a fast and efficient manner. However, given the simple polynomial expression we have, it may be used for all sorts of numerical computations such as computing derivatives, integrals, root finding, solving differential equations, amongst others.

> **Conclusion.** We have seen that the typical approximation methods (e.g. linear or spline interpolation, machine learning, Taylor) are suboptimal when approximating pricing functions inside Risk Engines because their convergence properties are generally poor. However, we have also seen that interpolation via Chebyshev polynomials has a very special property: exponential convergence. As a result, Chebyshev techniques are an ideal candidate for an accurate and computationally-efficient pricing approximation in a Risk Engine.

## Higher dimensions

The basic theory of Chebyshev polynomials and Chebyshev interpolants has been around for decades. However, extensions to higher dimensions are recent. First, we mention some of these extensions, particularly the ones we consider most relevant. Then we present the version that we find to be optimal in a practical setting.

Extensions to 2 and 3 dimensions are presented in (Townsend & Trefethen, 2013) and (Behnam & Trefethen, 2016), respectively. Both build on the 1-dimensional theory and implement their results in Chebfun, a Matlab package which was first developed for 1-dimensional functions using the theory of Chebyshev projections and interpolants.

The techniques presented in Chebfun are extremely powerful. The degree of approximation obtained is phenomenal. Moreover, derivatives of the functions are evaluated very efficiently among many other things. However, the techniques presented there are not ideal for many applications in finance as extreme accuracy (e.g. $10^{-15}$) is prioritised over speed of construction. In addition to this, it has the limitation of only being able to approximate functions of at most three dimensions.

More recently in (Gaß, Glau, Mahlstedt, & Mair, 2016), an extension of Chebyshev polynomials to $n$-dimensions was considered with particular attention to applications in finance. If the expression of a Chebyshev interpolant in one dimension is given by

$$p_n(x) = \sum_{k=0}^{n} c_k T_k(x),$$

the corresponding expression in higher dimensions is given by

$$p(\bar{x}) = \sum_{k_1=0}^{n_1} \ldots \sum_{k_d=0}^{n_d} c_{(k1,\ldots,kd)} T_{(k1,\ldots,kd)}(\bar{x}) \quad (4)$$

where $\bar{x} = (x_1, \ldots, x_d)$ is an element in $\mathbb{R}^d$,

$$T_{(k1,\ldots,kd)}(x_1, \ldots, x_d) = \prod_{i=1}^{d} T_{k_i}(x_i)$$

and the coefficients $c_{(k1,\ldots,kd)}$ are given by

$$\left(\prod_{i=1}^{d} \frac{2^{1(0<j_i<n_i)}}{n_i}\right) \sum_{k_1=0}^{n_1} {}''\ldots \sum_{k_d=0}^{n_d} {}'' Price\left(p^{(k_1,\ldots,k_d)}\right) \prod_{i=1}^{d} \cos\left(j_i \pi \frac{k_i}{n_i}\right), \qquad (5)$$

where " means the first and last summand are halved and $p^{(k_1,\ldots,k_d)}$ denotes a d-dimensional Chebyshev node.

In (Gaß, Glau, Mahlstedt, & Mair, 2016), Theorem 12 is extended to higher dimensions showing that under equivalent conditions to those in Theorem 12, we obtain sub-exponential convergence.

Also, practical applications are presented for 2-dimensional pricing functions, making the time savings evident while showing negligible precision loss when compared to standard bench-marks.

However, the expression in Equation 4 uses coefficients $c(k_1, \ldots, k_d)$ computed using Equation 5. This has the disadvantage that the number of operations grows exponentially with dimension. Therefore, as we increase dimensions, not only do we need to call the original pricing function an exponentially larger number of times, but we also suffer from the overhead of Equation 5 which grows exponentially too.

Our implementation uses an alternative technique that covers any given dimension. Instead of using the direct generalisation presented in Equation 4, we cover the domain of the function with 1-dimensional Chebyshev interpolants in order to produce the Chebyshev interpolant.

**Building the Interpolating Object.** The way it is built and evaluated is as follows. For purpose of illustration, let us focus on a two-dimensional function. An image of the domain is shown in the Figure below. The domain is defined by the interval $[x_1, x_1]$ along the $x$-axis and $[y_1, y_2]$ along the $y$-axis. A mesh of 2-dimensional Chebyshev points is obtained by first generating Chebyshev points along the $x$-axis, then along the $y$-axis, and then taking their Cartesian product to obtain the 2-dimensional Anchor point mesh, visible in solid blue dots in Figure 1. We call these points the interpolation "Anchor" points.

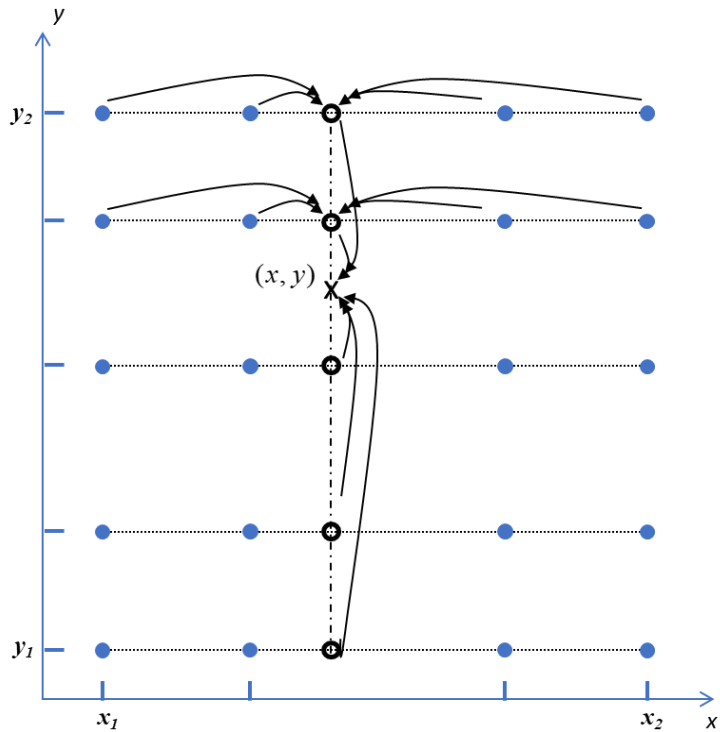

Once the mesh is constructed, one must evaluate the 2-dimensional function $f(x, y)$ at each Anchor point of the mesh. This is what we will call the "Building" phase.

**Evaluating with the Interpolating Object.** Now we describe the function evaluation at a generic point $(x, y)$. For each collection of Anchor points in the mesh running horizontally, consider a 1-dimensional Chebyshev interpolant. In the figure above, there are 5 of these interpolants each of which consists of 4 Anchor points. A central idea that we use here is that 1-dimensional Chebyshev methods are extremely accurate, so we can use a strictly-speaking "approximated" value as very close to the real value.

As a result, first, we use every 1-dimensional Chebyshev interpolant to obtain a value of $f(x, y)$ at each black circle on the diagram. Then, the collection of black circles is used to build a 1-dimensional Chebyshev interpolant, which in the figure above runs vertically. Note that the point $(x, y)$, where we want to evaluate the function, lies on the this new 1-dimensional interpolant. Therefore, we can evaluate this interpolant at $y$ to obtain the final value.

Observe that what we did in the first set of evaluations, when we computed the value of the horizontal 1-dimensional interpolants at the black circles, was to reduce the problem from 2 to 1 dimensions. By doing this we ended up with a single 1-dimensional interpolant which we can readily evaluate.

**Extension to any dimensionality.** The extension of this method to higher dimensions is straightforward. If we had a 3-dimensional function, instead of having a 1-dimensional family of 1-dimensional Chebyshev interpolants, as in the case described above, we would have a 2-dimensional family of 1-dimensional Chebyshev interpolants. By evaluating each of these interpolants, we reduce the dimension of the problem by one, leaving us with a 2-dimensional domain, identical in kind to the one we described above. If we extend this concept sequentially, we can extend Chebyshev methods to any dimension.

**Remarks.** Given the independence of each 1-dimensional Chebyshev interpolant in the described solution, if we know that a function is more difficult to approximate in some regions, we can tailor the grid construction so that the mesh has more resolution in those regions where we know that the function is more fluctuant. However, note that this is only an additional enhancement; we can always use a square or rectangular mesh for generic pricing functions and benefit from the exponentially convergence properties of Chebyshev techniques, which is the cornerstone of the power of these methods.

Furthermore, an additional advantage of using 1-dimensional Chebyshev interpolants in a multi-dimensional space is that their derivatives are straightforward to compute. That is particularly useful because the derivatives are given by the Chebyshev interpolants with the same computational cost that the function itself: we do not need to call the original function in more Anchor points (the part that is computationally expensive), and the same time it takes to evaluate the Chebyshev interpolant (negligible in most cases) is the same when we want to evaluate a derivative.

# Numerical Examples

The described methods can be applied in an industrial setting in the following way. The exhibit below shows the steps of a generic risk engine. Often, the Pricing step takes a vast amount of computational demand, either because we are calculating risk on a large number of vanilla trades, because of the presence of (difficult to price) exotic trades, or both.

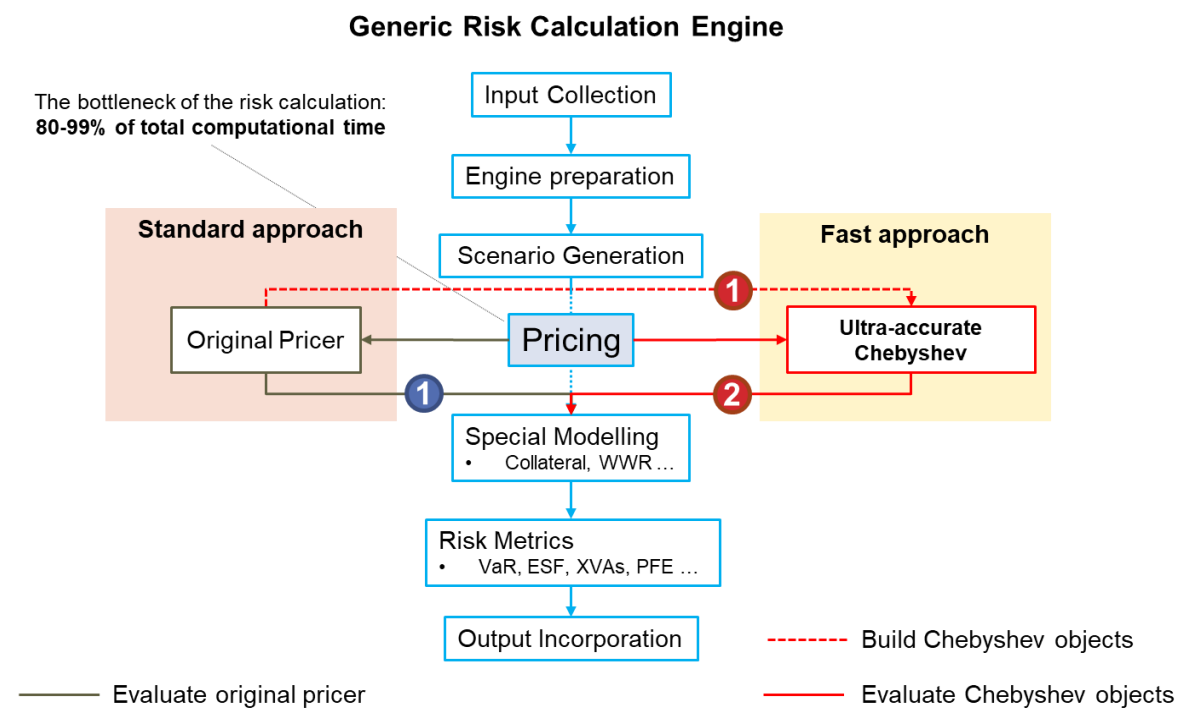

If the risk calculation is done in full-revaluation mode, the pricing step will call the Front Office "Original" pricing function many times, from several hundred up to a few million, depending on the calculation. This can be very costly from a computational standpoint. We call this setup the "brute-force" calculation.

The extraordinary properties of Chebyshev interpolants allow for a different optimised setup. Theorem 6 tells us that if we project the pricing function onto the space generated by the first $n$ Chebyshev polynomials, the projected function converges to the pricing function exponentially fast as $n$ increases. We also know that there are ways to compute the Chebyshev coefficients very effectively. For example, reference (Ahmed & Fisher, 1970) shows how we can compute those coefficients in $O(n \log(n))$ operations. We also know from Theorem 10 that we can have an ex-ante control on the error of the approximation. Finally, we know from Theorem 12 that the Chebyshev interpolant also converges exponentially, practically at the same rate as Chebyshev projections.

As a result, instead of the brute-force calculation we can build an ultra-accurate interpolating object using Chebyshev technology. This object is then invoked, instead of the original pricing function, to compute the portfolio price in each scenario.

This set up comprises two steps. First, in the "Building" phase, we call the Original pricing function a limited number of times to build an ultra-accurate Chebyshev interpolating object. By "Chebyshev interpolating object" we mean our code implementation of the explained Chebyshev interpolation framework discussed above, which can be generalised to any implementation that preserves the fundamentals explained. Then, in the second step, the "Evaluation" phase, the risk engine uses the optimised Chebyshev interpolating object to evaluate the portfolio prices in each scenario.

This optimisation method is good as long as (i) we call the original pricing function in the Building phase less times than in the brute-force approach, (ii) the precision of the interpolating object is very good and (iii) the computational cost of evaluating the interpolating object is low.

The exhibit below illustrates points (i) and (ii), compared to the typical linear interpolation technique (similar results are obtained with spline). The graph shows the maximum error of the interpolating objects with respect to the original pricing function as we increase the number of times the original pricing function is called to build the objects. This example is for a 2-dimensional Spot-Vol Surface in a Black-Scholes option.

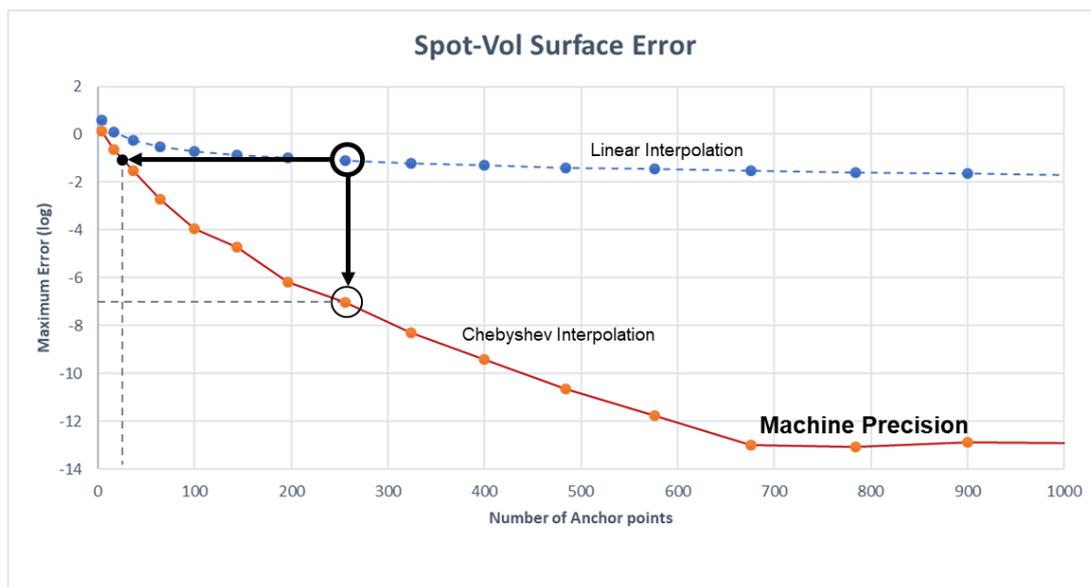

The blue dotted line is the maximum error of the linear interpolation scheme, the solid orange line is the maximum error of the Chebyshev scheme. Note the errors are given in a logarithmic scale. It can be clearly seen how the linear interpolation framework converges extremely slowly hence it lacks the precision needed in many applications. However, within the Chebyshev framework converges exponentially, so an extreme degree of accuracy is achieved with very little information from the pricing function. Indeed, machine precision can be achieved with only a few hundred points.

For example, let say that we are using a linear-interpolating scheme with around 250 Anchor points (a 16 x 16 mesh). The graph shows that we obtain an accuracy of around $10^{-1}$. If, on the one hand, we build a mesh with the same number of Anchor points (i.e. same computational effort in the Building phase) but apply the Chebyshev principles, the accuracy we get is of around $10^{-7}$; that is, full-revaluation accuracy in practice. If, on the other hand, we are happy with the accuracy, but we want to decrease the computational effort in the Building phase, the Chebyshev object achieves a $10^{-1}$ precision with only around 25 Anchor points (a 5 x 5 mesh); only 10% of the original computational cost.

The following table shows the improvement in pricing time, together with its accuracy, when using the Chebyshev objects comparted to a full-revaluation approach, using the QuantLib open source library. These are 1-dimenisional cases for illustrative purposes. The speed gains range from 100 in the case of already-considered fast analytic pricing functions, to nearly 6,000,000 for exotic products. The errors are mostly well below the basis-point precision, in spite of having trades like barriers, which have strong non-linear features. The only trade with an error outside the basis-point threshold is a trade that has been priced with a Monte Carlo simulation and which has an inbuilt error of the same magnitude; in this case the error comes from the lack of accuracy in the original pricing function itself, not the Chebyshev method.

| Pricing function | Pricing Method | QuantLib Run Time (ms) | Chebyshev Build Time (ms) | Chebyshev Run Time (ms) | Accuracy | Chebyshev Run Speed Multiplier |
|---|---|---|---|---|---|---|
| **IRS** | Analytic | 0.214 | 2.675 | 0.000103 | $10^{-15}$ | **2,088** |
| **European Option** | Analytic (BS) | 0.013 | 0.209 | 0.000127 | $10^{-6}$ | **110** |
| **American Option** | Monte Carlo | 23.103 | 247.117 | 0.000096 | $10^{-6}$ | **239,668** |
| **Bermudan Swaption** | Tree | 318.99 | 3642.12 | 0.000127 | $10^{-5}$ | **2,511,737** |
| **Barrier Option** | Analytic | 0.024 | 0.3 | 0.000125 | $10^{-4}$ | **192** |
| **Barrier Option** | Monte Carlo | 601.919 | 6590.522 | 0.000103 | $10^{-3}$ | **5,843,883** |

These illustrative examples are shown for low dimensional cases. These methods can be also applied to high-dimensional cases, as explained in the next section.

# Applications

The practical applications of these techniques are endless in risk management. Any risk computation that requires re-pricing a portfolio several times should benefit from it. Indeed, once the overhead of the Building phase is done, multiple calculations should benefit from the speed of the object built.

The following are practical examples of applications of these techniques in a commercial setting. The primary aim of this paper is to present the theoretical basis for the improvements that we see, so we are leaving detailed commercial explanations for subsequent publications.

## CVA, FVA and IMM capital of exotics

XVA and IMM Monte Carlo engines need to reprice portfolios many hundreds of thousands or even a few million times in each run. Full revaluation approaches can be very computationally expensive as soon as the portfolio of trades has non-vanilla products. Some institutions have implemented American Monte Carlo regression-based techniques. However, its implementation is far from trivial and the accuracy of the pricing proxy is far from ideal. As a result, regulators do not tend to like this solution for IMM computations. In addition, trying to get sensitivities out of it can also be troublesome.

Exotic pricing is a prime candidate to benefit from the Chebyshev techniques described. Indeed, we run an EPE and PFE-95% profile on a Bermudan Swaption in a two-factor model via brute-force revaluation with 1,000 scenarios and compare it to the same calculation via Chebyshev Objects.[4] The obtained profiles were exactly the same, as shown in the exhibit below, as the Chebyshev Objects gave perfect pricing accuracy in practice.

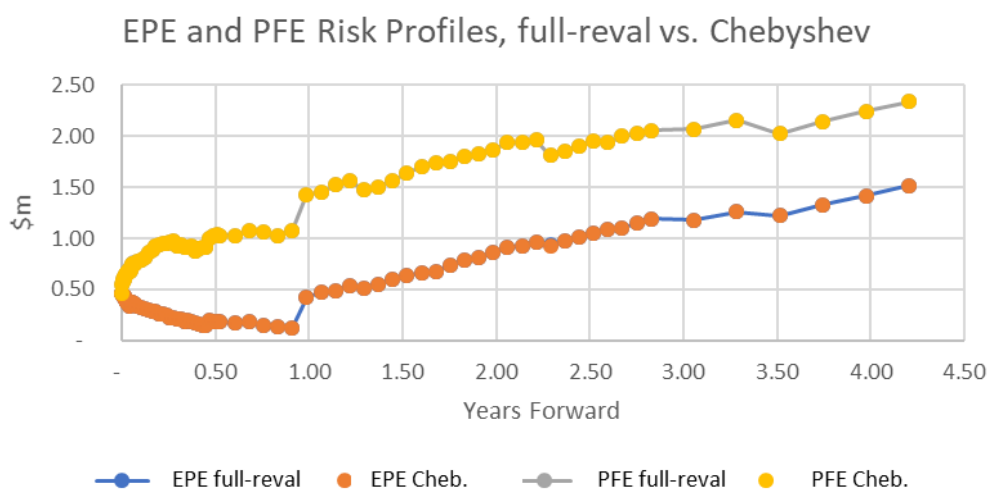

---

[4] Rates and volatility two-factor model.

The following table illustrates the computational improvement created.[5] The calculation via Chebyshev Objects dramatically decreases the computational effort compared to the benchmark full-revaluation. For example, a Monte Carlo simulation with 10,000 scenarios takes around 4.5 days in the single-core set up we use, while it only takes 53 minutes via Chebyshev Objects. When using the Chebyshev Objects, the incremental computational effort of increasing the number of scenarios is negligible; most of the effort in the proposed technique is taken by the Building step and, once that is finished, the computational effort to evaluate the interpolating objects is very small.

| Pricing Method | MC scenarios | Building step effort | Evaluation effort | Total effort | Chebyshev vs. full-reval |
|---|---|---|---|---|---|
| Full-reval | 1,000 | n/a | 10h:54m | 10h:54m | n/a |
| Full-reval | 10,000 | n/a | 4d:12h:58m | 4d:12h:58m | n/a |
| Full-reval | 100,000 | n/a | 45d:10h:05m | 45d:10h:05m | n/a |
| Chebyshev | 1,000 | 0h:53m | 70 milliseconds | 0h:53m:00s | **8%** |
| Chebyshev | 10,000 | 0h:53m | 700 milliseconds | 0h:53m:01s | **0.8%** |
| Chebyshev | 100,000 | 0h:53m | 7,000 milliseconds | 0h:53m:07s | **0.08%** |

This has multiple benefits. For example, not only can we perform risk profile computations speedily, but the speed of the evaluation time is so high that we can build the interpolating object, save it and reuse as often as needed for many ultra-fast XVA sensitivity calculations via bump-and-reval.

A remarkable application is that, once the interpolating object has been built, we can easily simulate sensitivities, as shown in the next section.

### Initial Margin Simulation

We have seen that the Chebyshev Objects easily achieve an accuracy of, say, $10^{-6}$. We achieve that accuracy with a function

$$P(x) \cong p_n(x) = \sum_{k=0}^{n} c_k T_k(x).$$

Given that Chebyshev polynomials are very well understood, their derivatives can be computed very easily. Consequently, given the high accuracy of the approximation, the derivative of the Pricing function $P(x)$ with respect to its risk factors can be approximated by

$$P'(x) \cong p_n'(x) = \sum_{k=0}^{n} c_k T_k'(x).$$

Strictly speaking, this derivative approximation tends to lose a bit of accuracy relative to the approximation of the original function. However, the Chebyshev interpolant tends to be such a good approximation to the pricing function that the derivative of the interpolator tends also to be a very good proxy for the derivative.

---

[5] Computation in C++, using QuantLib pricing functions, in a desktop PC Intel i7, single core process. Full-revaluation computations done for 1,000 scenarios.

This has a very important consequence: the Chebyshev approximation to the pricing function not only enables efficient full-revaluation in a risk engine, it also provides a very accurate and very efficient simulation of the sensitivities without any extra Building computational cost. Risk simulations of Initial Margins, or hedging simulations, are surprisingly easy in this framework.

The following graphs shows the expected profile, $5^{th}$ and $95^{th}$ percentile of a SIMM™ simulation using this methodology for a portfolio of swaps, Bermudan swaptions and barrier equity options. The simulation was done in no-time and with absolute precision. The Monte Carlo engine simulated around 50 million sensitivities (36 million deltas and 14 million Vegas) in 2.3 seconds in a standard laptop via Chebyshev interpolants of the pricing functions.[6] The accuracy of the sensitivities was better than $10^{-4}$; i.e. perfect simulation in practice.

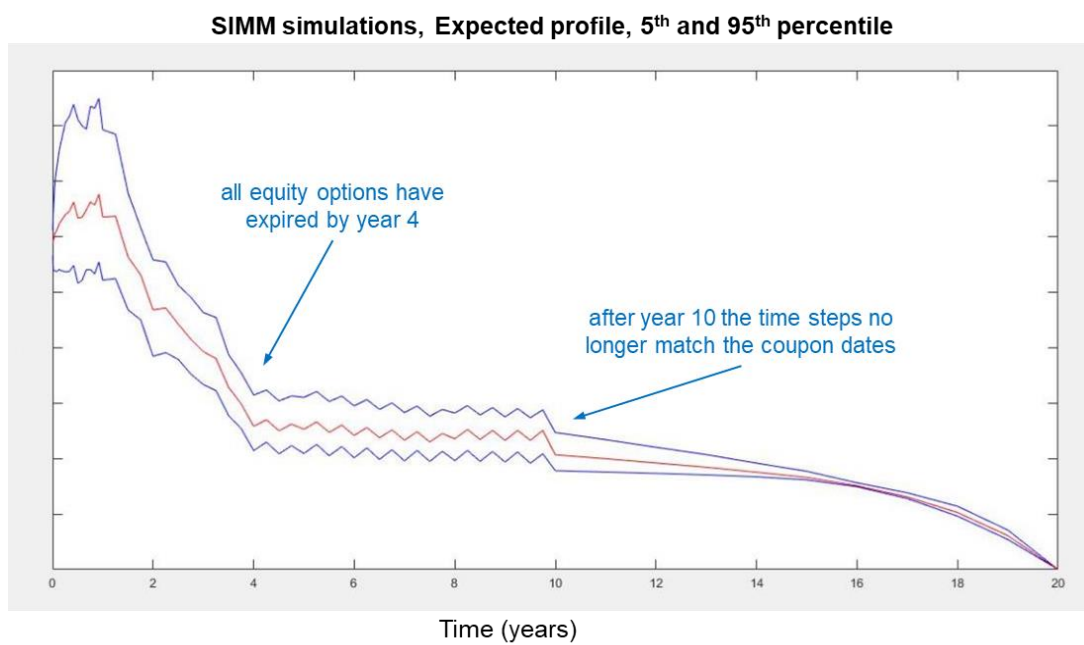

## IMA-FRTB

Another application sits in the Internal Models Approach for FRTB (IMA-FRTB). The technique we explain here has been tested in a commercial tier-1 bank in that context; a publication will be available soon. The technique was applied to a portfolio of swaps, swaptions and Bermudan swaptions.

In this case, the technique had the extra challenge of the dimensionality of the input space, which could be of up to several hundred dimensions if we include all interest rate spread curves and volatility surfaces. A dimensionality reduction technique was applied, comprising of a blend of PCA and orthogonal slide aggregation; the details of such technique are beyond the scope of this paper, they will be published soon in a subsequent article.

The following graph illustrates the 1day-PnL correlation between the original pricing function and the fast pricing function after applying this technique to a Bermudan swaption (left pane), as well as the capital charge (based on 10day expected shortfall) obtained through both techniques (right pane). It can be seen how the error is so small that we can be sure to pass the PnL attribution tests; indeed, the PnL correlation was of 99.9%, ensuring P&L Attribution Test compliance. Importantly, the computing load needed in the technique that makes use of the approximating technique was between 1% and 5% of that needed in a brute-force full-revaluation. This technique can save a bank many millions in hardware cost annually.

---

[6] Evaluation times. Building times are considered to happen in an overnight calculation batch.

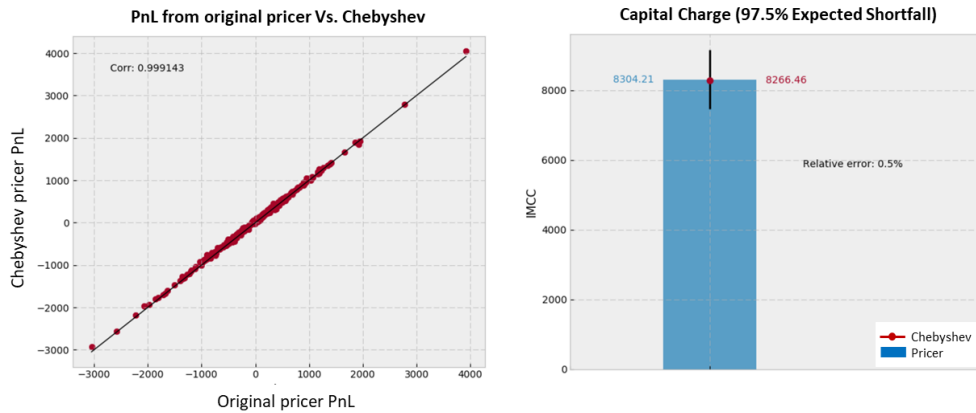

### Pricing function "Cloning"

Often, risk engines live in separate IT infrastructures to the pricing engines. One of the central problems of risk engines is how to transport a pricing function that lives, typically, in the Front Office systems, into the simulation framework. The Chebyshev framework offers a solution for it.

For example, let's say that the risk engine needs to price a product 3,000 times. The Chebyshev Objects only need to price each product on very few Anchor points in order to replicate the pricing function very accurately. In a typical example we only need the value of the pricing function on 150 Anchor points. The Risk Engine can set those 150 Anchor points as the Chebyshev "calibration" scenarios, query the price of the product only on those 150 points, and replicate the pricing function inside the Risk Engine with very high precision and ultra-fast evaluation time.

Note that this communication between the Risk Engine and the Pricing Engine can be done dynamically or statically. In a dynamic mode, the Risk Engine has a handle to the pricing routine that sits in the Pricing Engine, so it can query a pricing job "on-the-fly" whenever it is needed. However, many systems do not have that capability, in which case the 150 pricing scenarios can be communicated to the Pricing Engine in one single batch, and the 150 prices can also be communicated back to the Risk Engine in another single batch. Indeed, this communication can be as manual as a flat file or Excel spreadsheet, because the Anchor points are given by the Chebyshev points, and so a dynamic communication with the Pricing Engine, or specific information of the trade, is not needed. In many practical contexts, this is a key feature of this methodology.

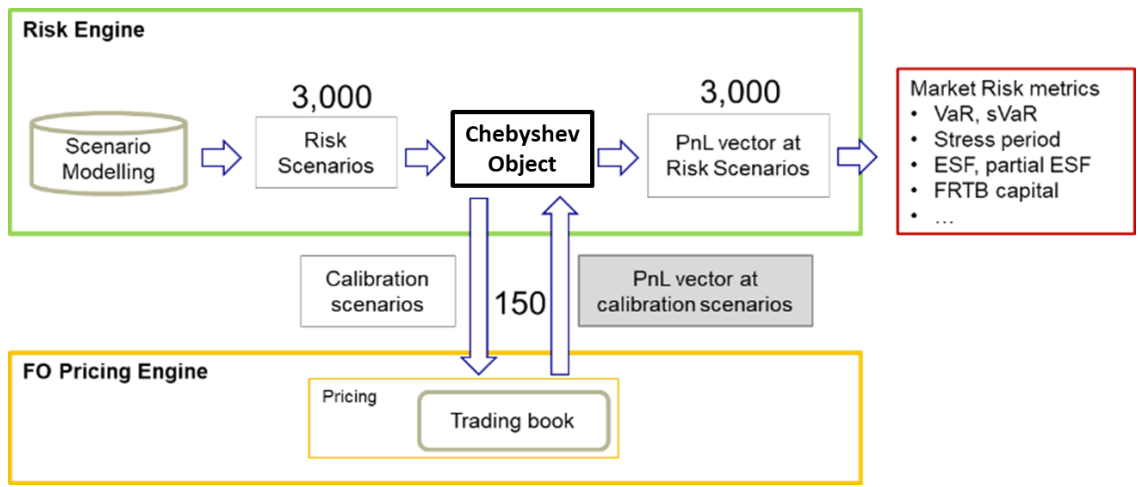

## Portfolio Pricing Compression

A further computational improvement can be achieved when we have portfolios of trades on the same risk factors. For example, let's say that we have 5,000 trades that are driven by the same family of risk factors (e.g. swaps and options on USD yield curves and volatility surfaces). Each of its pricing functions can be represented by a Chebyshev interpolant, as we have seen. We have, in this example, 5,000 single-trade Chebyshev Objects.

Trade 1     $P_1(x) = \sum_{k=0}^{n} c_k^1 T_k(x)$

Trade 2     $P_2(x) = \sum_{k=0}^{n} c_k^2 T_k(x)$

...

Trade 5,000     $P_{5000}(x) = \sum_{k=0}^{n} c_k^{5000} T_k(x)$

Portfolio

$P_{1+2+\cdots+5000}(x) = \sum_{k=0}^{n} \left(c_k^1 + c_k^2 + \cdots + c_k^{5000}\right) T_k(x)$

In order to run multiple revaluations of the portfolio, instead of evaluating the 5,000 Chebyshev Objects, we could decrease further the computing effort by adding up the Chebyshev coefficients and constructing, in this way, a *portfolio* Chebyshev Object. The computational effort of this new Object is the same of a single-trade Object, hence we achieve a further 5,000 times acceleration in the revaluation effort.

## AAD

Needless to say, Adjoint Algorithmic Differentiation (AAD) is a fantastic state-of-the-art methodology that can enable the computation of many sensitivities with limited computational effort. As it is well known in the industry, it presents a challenge of implementation and maintenance. For this reason, several trading houses have decided to postpone or avoid its implementation.

A central problem is that, in order to implement AAD, the developers need to implement and maintain the adjoint version of each pricing function. Also, when running the simulation, memory is an issue, as the computation needs to store an enormous amount of information.

Again, Chebyshev may be used to reduce the complexity of these problems. If we have the pricing functions of a number of trades represented by a Chebyshev interpolant, we can code up a generic adjoint version for all of them, hence considerably reducing implementation and maintenance effort, as well as memory footprint.

$$P(x) = \sum_{k=0}^{n} c_k T_k(x) \quad \Longrightarrow \quad \overline{P(x)} = \overline{\sum_{k=0}^{n} c_k T_k(x)}$$

It must be noted that this AAD generic method is only a theoretical idea for now. We have not had the chance to test it. Please contact us if you are an AAD expert and would like to collaborate with us in this subject.

### Other Applications

As indicated, the number of applications of Chebyshev interpolants and Chebyshev Objects is endless in the context of risk calculations. In addition to those shown, other clear candidates include stress testing and CCAR programs, hedging simulations, model risk management, etc.

## Comparison with Other Methods

Chebyshev interpolants may "compete" with the following alternatives

- **Full-revaluation.** Obviously, if full-revaluation is a practical option, there is no need for any algorithmic alternative. However, the financial industry is now suffering from its high computational demand. Indeed, that is why it is often referred to as the "brute-force" solution, as the only way to overcome its computation limitation is by adding more hardware.
- **Standard Interpolations.** These methods include the popular linear or cubic-spline interpolation methods. As it has been clearly proven in this paper, Chebyshev interpolation outperforms any of those methods, in all counts, by orders of magnitude.
- **Regression-based "Longstaff-Schwartz" approximations.** Another somewhat popular alternative to full-revaluation in risk calculations has been the adaptation of methods used to price options via American Monte Carlo (Cesari, et al., 2009). Before we knew of the power of Chebyshev methods, regression methods seemed to be, often, the only possibility for efficient computations. However, it is well known that they suffer from a very significant regression error (that the user *hopes* to have cancelled out on average, but with limited control over it), very poor performance in the tails of the risk distribution (hence risk departments and regulators tend to disregard it) and the calibration of the regression tends to be more of a "black art" than a scientific method. In contrast, Chebyshev interpolation methods solve all these problems and require similar computational efforts.

## Independent Validation

Reference (Gaß, Glau, Mahlstedt, & Mair, 2016) constitutes an independent validation of the methods described in this paper. The authors of that 2016 paper, M. Gaß, K. Glau, M. Mahlstedt and M. Mair, describe mostly the same methodology that we outline in this paper.

Indeed, (Gaß, Glau, Mahlstedt, & Mair, 2016) states that "*for pricing tasks in mathematical finance Chebyshev interpolation still seems to be rarely used and its potential is yet to be unfolded*". It also states that "*The financial implications of the high precision achieved with such a small number of interpolation nodes are twofold. First, it shows that there are interesting cases in which we observe an accuracy in the range of machine precision. In this comfortable situation without methodological risk we can ignore the fact that approximations are implemented. Second, compared with other sources of risk, already errors from a much lesser accuracy level can be ignored. If we agree that an accuracy of $10^{-4}$ is satisfactory, already 36–49 interpolation nodes for the approximation of call option prices […] are sufficient*"

## Bibliography

Ahmed, N., & Fisher, P. (1970). Study of algorithmic properties of Chebyshev coefficients. *Int. J. Computer Math*, 307–317.

Behnam, H., & Trefethen, L. (2016). Chebfun in three dimensions. *Preprint*.

Bernstein, S. (1912). *Sur l'Ordre de la Meilleure Approximation des Fonctions Continues par des Polynomes de Degré Donné.* Mém. Acad. Roy. Belg.

Cesari, G., Aquilina, J., Charpillon, N., Filipovic, Z., Lee, G., & Manda, I. (2009). *Modelling, Pricing and Hedging Counterparty Credit Exposure: A Technical Guide.* Springer.

Clenshaw, C., & Curtis, A. (1960). A method for numerical integration on an automatic computer. *Numer. Math. 2*, 197–205.

Ehlich, H., & Zeller, K. (1966). Auswertung der Normen von Interpolationsoperatoren. *Math. Ann. 164*, 105–112.

Faber, G. (1914). Über die interpolatorische Darstellung stetiger Funktionen. *Jahresber. Deutsch. Math. Verein 23*, 192–210.

Gaß, M., Glau, K., Mahlstedt, M., & Mair, M. (2016). Chebyshev Interpolation for Parametric Option Pricing. *Preprint. https://arxiv.org/abs/1505.04648*.

Handscomb, D., & Mason, J. (2003). *Chebyshev Polynomials.* Chapman and Hall/CRC.

Runge, C. (1901). Über empirische Funktionen and die Interpolation zwischen äquidistanten Ordinaten. *Z. Math. Phys.*, 224–243.

Townsend, A., & Trefethen, L. (2013). An Extension of Chebfun to Two Dimensions. *SIAM Journal on Scientific Computing 35 (6)*, C495–C518.

Trefethen, L. (2013). *Approximation Theory and Approximation Practice.* SIAM.

Weierstrass, K. (1885). Über die analytische Darstellbarkeit sogenannter willkürlicher Functionen einer reellen Veranderlichen. *Sitzungsberichte der Akademie zu Berlin* , 633–639.